# Hyperbolic-Type Orbits in the Schwarzschild Metric

F.T. Hioe* and David Kuebel

Department of Physics, St. John Fisher College, Rochester, NY 14618

and

Department of Physics & Astronomy, University of Rochester, Rochester, NY 14627, USA

August 11, 2010

**Abstract**

Exact analytic expressions for various characteristics of the hyperbolic-type orbits of a particle in the Schwarzschild geometry are presented. A useful simple approximation formula is given for the case when the deviation from the Newtonian hyperbolic path is very small.

## 1 Introduction

In a recent paper [1], we presented explicit analytic expressions for all possible trajectories of a particle with a non-zero mass and for light in the Schwarzschild metric. They include bound, unbound and terminating orbits. We showed that all possible trajectories of a non-zero mass particle can be conveniently characterized and placed on a "map" in a dimensionless parameter space $(e, s)$, where we have called $e$ the energy parameter and $s$ the field parameter. One region in our parameter space $(e, s)$ for which we did not present extensive data and analysis is the region for $e > 1$. The case for $e > 1$ includes unbound orbits that we call hyperbolic-type as well as terminating orbits. In this paper, we consider the hyperbolic-type orbits in greater detail.

In Section II, we first recall the explicit analytic solution for the elliptic- ($0 \leq e < 1$), parabolic- ($e = 1$), and hyperbolic-type ($e > 1$) orbits in what we called Region I ($0 \leq s \leq s_1$), where the upper boundary of Region I characterized by $s_1$ varies from $s_1 = 0.272166$ for $e = 0$ to $s_1 = 0.250000$ for $e = 1$ to $s_1 \to 0$ as $e \to \infty$. The analytic expression for the orbits is given in terms of the Jacobian elliptic functions of modulus $k$.

In Section III, we consider the hyperbolic-type orbits specifically and present tables and curves of constant $k^2$ for $e > 1$; the corresponding tables and curves for $0 \leq e \leq 1$ in Region I were given in ref.1. We also present examples of these orbits for several specific values of $s$ and a given value of $e$, as $s$ varies from $s \simeq 0$ and $k^2 \simeq 0$ corresponding to the Newtonian hyperbolic orbit to $s = s_1$ corresponding to what we call the asymptotic hyperbolic-type orbit. The two asymptotes of the hyperbolic-type orbit are of particular interest. Their angles with respect to and their points of intersections on the horizontal axis are presented by exact analytic expressions.

In Section IV, we consider the case when the hyperbolic-type orbit deviates only slightly from the Newtonian hyperbolic orbit. We present a simple approximation formula that gives the deviation of the asymptotes from the asymptotes of the Newtonian hyperbola. The approximate formula is analogous to but not the same as that for the well known precession of the perihelion for an elliptic-type orbit. The derivation of this approximate formula reveals some extra caution that needs to be taken for the calculations to be correct, and the additional steps that are necessary may explain why this result has not been presented earlier. In the papers that discussed the deflection of spacecraft trajectories as a test of general relativity, Longuski et al [2] derived and used a deflection formula that they stated to be valid for the parametrized post-Newtonian metric that includes the Schwarzschild metric as a special case. We use the same input data that Longuski et al used but instead apply our approximate formula to calculate the deviation for the trajectories of the spacecraft, and find slightly larger values than what they reported.

In Section V, we briefly mention the terminating orbit in Region I for the case $e > 1$ for completeness, as it is basically similar to that for the case $0 \leq e \leq 1$ that has been discussed in ref.1.

In Section VI, we give a summary of our results. A brief discussion of how our analytic expressions for various characteristics for the hyperbolic-type orbits reduce to the Newtonian case is presented in Appendix A, and a special case for $k^2 = 1/2$ is presented in Appendix B.

## 2 Analytic Expression for the Orbits, Bound and Unbound

We begin with the analytic expression for the orbit, its essential features, and the parameters for describing it, that were given by us [1] recently. We consider the Schwarzschild geometry, i.e. the static spherically symmetric gravitational field in the empty space surrounding some massive spherical object (which we shall refer to as a star or a blackhole even though this also includes the case of a planet such as Jupiter or Earth) of mass $M$, and we consider a particle (which can be a spacecraft, an asteroid, or a planet), with its position relative to the star or blackhole described in the spherical coordinates $r, \theta, \phi$, moving in the equatorial plane $\theta = \pi/2$. We define three parameters $\kappa$, $h$ and $\alpha$ by

$$\kappa \equiv \frac{E}{m_0 c^2}, \tag{1}$$

$$h \equiv r^2 \dot{\phi}, \tag{2}$$

and

$$\alpha \equiv \frac{2GM}{c^2}, \tag{3}$$

where $E$ is the total energy of the particle in its orbit and $m_0$ is its rest mass, and $h$ can be identified as the angular momentum per unit rest mass of the particle. The derivative $\cdot$ represents $d/d\tau$ where $\tau$ is the proper time along the path, and $\alpha$ is the Schwarzschild radius, where $G$ is the universal gravitation constant, and $c$ is the speed of light. It is convenient to define a dimensionless parameter $U$

$$U = \frac{1}{4}\left(\frac{\alpha}{r} - \frac{1}{3}\right). \tag{4}$$

Using eq.(2), the 'combined' energy equation [3]

$$\dot{r}^2 + \frac{h^2}{r^2}\left(1 - \frac{\alpha}{r}\right) - \frac{c^2\alpha}{r} = c^2(\kappa^2 - 1) \tag{5}$$

can be expressed as

$$\left(\frac{dU}{d\phi}\right)^2 = 4U^3 - g_2 U - g_3. \tag{6}$$

Here the coefficients $g_2$ and $g_3$ are given by

$$\begin{aligned} g_2 &= \frac{1}{12} - s^2, \\ g_3 &= \frac{1}{216} - \frac{1}{12}s^2 + \frac{1}{4}(1 - e^2)s^4, \end{aligned} \tag{7}$$

where the dimensionless parameters $e$ and $s$ are defined by

$$e = \left[1 + \frac{h^2 c^2(\kappa^2 - 1)}{(GM)^2}\right]^{1/2}, \tag{8}$$

and

$$s = \frac{GM}{hc}. \tag{9}$$

We call $e$ the energy eccentricity parameter or simply the energy parameter, and we call $s$ the gravitational field parameter or simply the field parameter. Both the energy parameter $e$ and the field parameter $s$ have values that range from 0 to $\infty$. Characterizing all possible orbits in the parameter space $(e, s)$ is one of the important results we presented in ref.1. Note also the distinction that we made between the energy eccentricity parameter $e$ and the true eccentricity of a bound orbit [1].

The discriminant $\Delta$ defined by

$$\Delta = 27g_3^2 - g_2^3 \tag{10}$$

of the cubic equation

$$4U^3 - g_2 U - g_3 = 0 \tag{11}$$

divides the parameter space $(e, s)$ into two regions: Region I corresponding to $\Delta \leq 0$ covers the values of the field parameter $0 \leq s \leq s_1$, and Region II corresponding to $\Delta > 0$ covers the values of $s > s_1$, where $s_1$ as a function of $e$ is given by

$$s_1^2 = \frac{1 - 9e^2 + \sqrt{(1 - 9e^2)^2 + 27e^2(1 - e^2)^2}}{27(1 - e^2)^2}, \tag{12}$$

for $e \neq 1$, and $s_1^2 = 1/16$ for $e = 1$. Region I has terminating and non-terminating orbits that include bound (for $0 \leq e < 1$) and unbounded orbits (for $e \geq 1$), and Region II has terminating orbits only [1].

In Region I ($\Delta \leq 0$), the three roots of the cubic equation (11) are all real. We call the three roots $e_1, e_2, e_3$ and arrange them so that $e_1 > e_2 > e_3$. Define a dimensionless distance $q$ measured in units of the Schwarzschild radius $\alpha$ by

$$q \equiv \frac{r}{\alpha}, \tag{13}$$

which is related to $U$ of eq.(4) by

$$\frac{1}{q} = \frac{1}{3} + 4U. \tag{14}$$

The equation for the non-terminating orbits in Region I, bound and unbound, and for $0 \leq e \leq \infty$, is given [1] in terms of Jacobian elliptic functions [4] of modulus $k$ by

$$\frac{1}{q} = \frac{1}{3} + 4e_3 + 4(e_2 - e_3)sn^2(\gamma\phi, k) \tag{15}$$

$$= \frac{1}{3} + 4e_3 + 4(e_2 - e_3)\frac{1 - cn(2\gamma\phi, k)}{1 + dn(2\gamma\phi, k)}, \tag{16}$$

where $e_1 > e_2 \geq U > e_3$. Equation (15) gives the orbit of a particle in the gravitational field of a star or blackhole of mass $M$ that is described by a two-dimensional polar coordinate $(q, \phi)$ in terms of its dimensionless distance $q$ and polar angle $\phi$ relative to the star that is situated at the origin [5]. The constant $\gamma$ appearing in the argument, and the modulus $k$, of the Jacobian elliptic functions are given in terms of the three roots of the cubic equation (11) by

$$\gamma = (e_1 - e_3)^{1/2}, \tag{17}$$
$$k^2 = \frac{e_2 - e_3}{e_1 - e_3}. \tag{18}$$

where $e_1, e_2, e_3$ are given by

$$\begin{aligned} e_1 &= 2\left(\frac{g_2}{12}\right)^{1/2} \cos\left(\frac{\theta}{3}\right), \\ e_2 &= 2\left(\frac{g_2}{12}\right)^{1/2} \cos\left(\frac{\theta}{3} + \frac{4\pi}{3}\right), \\ e_3 &= 2\left(\frac{g_2}{12}\right)^{1/2} \cos\left(\frac{\theta}{3} + \frac{2\pi}{3}\right), \end{aligned} \qquad (19)$$

and where

$$\cos\theta = g_3 \left(\frac{27}{g_2^3}\right)^{1/2}. \qquad (20)$$

The modulus $k$ of the elliptic functions has a range $0 \leq k^2 \leq 1$. The period of $cn(2\gamma\phi, k)$ is $4K(k)$, and the period of $dn(2\gamma\phi, k)$ and of $sn^2(\gamma\phi, k)$ is $2K(k)$, where $K(k)$ is the complete elliptic integral of the first kind [4]. For $k = 0$, $sn(x,0) = \sin x$, $cn(x,0) = \cos x$, $dn(x,0) = 1$, and for the case of $k^2 = 1$, $sn(\gamma\phi, 1) = \tanh(\gamma\phi)$, $cn(\gamma\phi, 1) = dn(\gamma\phi, 1) = \sec h(\gamma\phi)$. As $k^2$ increases from 0 to 1, $K(k)$ increases from $\pi/2$ to $\infty$.

The orbit equation (15) has been obtained with the following boundary or initial condition: For $0 \leq e \leq \infty$, the point at $\phi = K(k)/\gamma$ has been chosen to give $U = e_2$ or $q = q_{\min}$ (see eq.22 below). The point at $\phi = 0$ gives $U = e_3$ or $q = q_{\max}$ for the case $0 \leq e \leq 1$ but gives no real value for $U$ or $q$ for the case $e > 1$ for reasons that become clear in Section III.

The orbits given by eq.(15) for $0 \leq e < 1$, $e = 1$, and $e > 1$ will be referred to as the elliptic-, parabolic- and hyperbolic-type orbits respectively, even though their shapes can differ greatly from the conic sections that arise in Newtonian mechanics [1]. They correspond to the cases for which $0 \leq \kappa^2 < 1$, $\kappa^2 = 1$, and $\kappa^2 > 1$ respectively, where $\kappa$ is defined by eq.(1).

The properties of the elliptic-type and parabolic-type orbits are described in detail in ref.1. For the purposes of this paper, we note the expression for the precession angle

$$\Delta\phi = \frac{2K(k)}{\gamma} - 2\pi. \qquad (21)$$

which we will need when discussing the perturbative results in Section IV. We also note the expression for the minimum distance given by

$$\frac{1}{q_{\min}} = \frac{1}{3} + 4e_2. \qquad (22)$$

which we will use to calculate the distance of closest approach for the hyperbolic-type orbits.

In the Newtonian limit, $s \simeq 0$ and $k^2 \simeq 0$. The constant $c^2(\kappa^2 - 1)$ which is $< 0$ for a bound orbit and $\geq 0$ for an unbounded orbit, can be identified with $2E_0/m$ in the Newtonian limit, where $E_0$ is the sum of the kinetic and potential energies and is $< 0$ for a bound orbit, and is $\geq 0$ for an unbound orbit, and $m$ is the mass of the particle (which approaches $m_0$), and that

$$e \simeq \left[1 + \frac{2E_0 h^2}{m(GM)^2}\right]^{1/2}, \tag{23}$$

and eq.(15) reduces to an orbit equation given by

$$\frac{1}{r} \simeq \frac{GM}{h^2}\left\{1 - e\left[1 - 2sn^2(\gamma\phi, k)\right]\right\}$$

that represents a precessing elliptical orbit, and to the Newtonian orbit given by

$$\frac{1}{r} \simeq \frac{GM}{h^2}(1 - e\cos\phi) \tag{24}$$

for $e > 0$. The case for $e = 0$ is special and interesting in its own right, and will be described in another paper [6].

We note that, just like the Newtonian orbit represented by eq.(24) for which the cases $0 \leq e < 1$ and $e \geq 1$ give rise to bound (elliptic) and unbound (parabolic and hyperbolic) orbits, the single analytic expression represented by eq.(15) with the same criteria for $e$ gives rise to precessing bound (elliptic-type) and unbound (parabolic- and hyperbolic-type) orbits, even though there is a greater variety of behavior for each type. We should also note that in general relativity a second parameter $s$ with a range $0 \leq s \leq s_1$, where $s_1$ is given by eq.(12), is needed for these three types of orbits to be realized [1]. For the case $s > s_1$, all orbits become terminating.

## 3 Hyperbolic-Type Orbits

We now consider the hyperbolic-type orbits given by eq.(15) when $e > 1$. As was shown in ref.1, the range of values of $s$ for which these hyperbolic-type orbits exist is $0 \leq s \leq s_1$, where $s_1$ is given by eq.(12), and it ranges from $s_1 = (2/27)^{1/2} = 0.272166$ for $e = 0$ to $s_1 = 1/4 = 0.250000$ for $e = 1$, and then it decreases more quickly as $e$ increases beyond 1 and $s_1^2 \to (\sqrt{27}e)^{-1} \to 0$ as $e \to \infty$ [7]. Also derived in ref.1 is a direct relationship between $k^2$ given by eq.(18) and $e$ and $s$ given by eqs.(8) and (9):

$$\frac{1 - 18s^2 + 54(1-e^2)s^4}{(1 - 12s^2)^{3/2}} = \frac{(2-k^2)(1+k^2)(1-2k^2)}{2(1-k^2+k^4)^{3/2}}.$$

We use this equation to give a plot of lines of constant $k^2 = 0.001, 0.01, 0.1, 0.3, 0.7, 1$ as shown in Fig.1. The values of $s$ on these constant $k^2$ lines for the values of $e = 1, 2, 3, ..., 10$ are given in Table I, which thus give the coordinates $(e, s)$ of the points on the lines representing different values of $k^2$. These coordinate points $(e, s)$ from Table I are to be used with Tables II-VII which will present various characteristics of the hyperbolic-type orbits at the corresponding coordinate points.

The smallest root $e_3$ given by eq.(19) is less than $-1/12$ for $e > 1$, and from eq.(15), $q$ becomes infinite when the polar angle $\phi = \Psi_1$, where $\Psi_1$ is given by

$$sn(\gamma \Psi_1, k) = \sqrt{-\frac{\frac{1}{3} + 4e_3}{4(e_2 - e_3)}}, \tag{25}$$

or

$$\Psi_1 = \gamma^{-1} sn^{-1} \left( \sqrt{-\frac{\frac{1}{3} + 4e_3}{4(e_2 - e_3)}}, k \right),$$

and where $\gamma$ and $k$ are defined by eqs.(16) and (17). As $\phi$ increases from $\Psi_1$,

$$\Psi_2 \equiv \frac{2K(k)}{\gamma} - \Psi_1 \tag{26}$$

is the next value of $\phi$ for $q$ to become infinite. Thus eq.(15) gives the hyperbolic-type orbit for

$$\Psi_1 \leq \phi \leq \Psi_2 \tag{27}$$

in Region I. The orbit given by eq.(15) describes a particle coming from infinity along an incoming asymptote at an angle $\Psi_1$ to the horizontal axis, turning counter-clockwise about the star to its right on the horizontal axis, and leaving along an outgoing asymptote at an angle $\Psi_2$. The minimum distance $q_{\min}$ of the particle from the star or blackhole is given from eq.(15) when the particle is at a polar angle $\phi = K(k)/\gamma$, where $dr/d\phi = 0$ at $q = q_{\min}$ given by eq.(22). A straight line through the origin $O$ (where the star or blackhole is located) that makes an angle

$$\chi = \frac{K(k)}{\gamma} \tag{28}$$

with the horizontal axis is the symmetry axis of the hyperbolic-type orbit. We note that the particle intersects the horizontal axis at a distance $q_h$ from the star or blackhole which can be obtained from eq.(15) by setting the polar angle $\phi = \pi$ and that $q_h$ is not equal to $q_{\min}$ generally, except in the Newtonian limit.

We next find where the two asymptotes intersect the horizontal axis. Let $E = mc^2 = m_0 c^2 / \sqrt{1 - v_\infty^2/c^2}$ be the total energy of the particle of rest mass $m_0$ initially very far away from the star or blackhole where $v_\infty$ is the initial velocity of the particle, and let $b$ be the impact parameter, i.e. the perpendicular distance from the origin (where the star or blackhole is located) to the asymptote. We can express $h = mv_\infty b/m_0$, the angular momentum of the particle per unit rest mass of the particle as

$$h = \frac{v_\infty b}{\sqrt{1 - v_\infty^2/c^2}}. \tag{29}$$

Denoting $\beta \equiv v_\infty/c$, and from eqs.(29) and (9), the dimensionless impact parameter $b/\alpha$, where $\alpha$ is the Schwarzschild radius given by eq.(3), can be expressed as

$$\frac{b}{\alpha} = \frac{\sqrt{1-\beta^2}}{2s\beta}. \tag{30}$$

From eqs.(8), (1), and (9), the energy parameter $e$ can be expressed as

$$e = \sqrt{1 + \frac{1}{s^2} \frac{\beta^2}{1-\beta^2}}. \tag{31}$$

From eq.(31), we can write

$$e^2 - 1 = \frac{1}{s^2} \frac{\beta^2}{1-\beta^2}. \tag{32}$$

Equation (32) can be used with eq.(30) to show

$$\frac{b}{\alpha} = \frac{1}{2s^2\sqrt{e^2-1}}. \tag{33}$$

Since the asymptotes make an angle $\Psi_1$ and $\Psi_2$ respectively with the horizontal axis, their intersection points on the horizontal axis are at a distance $r_{a1}$ and $r_{a2}$, or at a dimensionless distance $q_{a1} \equiv r_{a1}/\alpha$ and $q_{a2} = r_{a2}/\alpha$ respectively from the origin that are given by

$$q_{ai} = \left| \frac{b}{\alpha} \frac{1}{\sin \Psi_i} \right|, i = 1, 2,$$

where $\Psi_1$ and $\Psi_2$ are given by eqs.(25) and (26). The $x$ coordinates of the points where the two asymptotes intersect the horizontal axis relative to the origin are given by

$$\begin{aligned} x_{a1} &= -\frac{1}{2s^2\sqrt{e^2-1}} \frac{1}{\sin \Psi_1}, \\ x_{a2} &= \frac{1}{2s^2\sqrt{e^2-1}} \frac{1}{\sin \Psi_2}. \end{aligned} \tag{34}$$

A negative (positive) sign for the $x_{ai}$ means that the corresponding asymptote intersects the horizontal axis to the left (right) of the origin.

The symmetry axis of the hyperbolic-type orbit, which is a straight line through the origin that makes an angle $\chi$ with the horizontal axis given by eq.(28), passes through the point where the hyperbolic-type orbit is at a minimum distance from the origin, and it also passes through the intersection point of the two asymptotes. This symmetry axis coincides with the horizontal axis in the Newtonian limit. When Cartesian coordinates are used to describe a Newtonian hyperbolic orbit, the point of intersection of the symmetry axis of the hyperbola with the two asymptotes is usually taken to be the origin of the coordinate system, in contrast to the point which is taken to be the origin of the polar coordinate system that we use. We give a brief discussion in Appendix A on how the general relativistic hyperbolic-type orbit and its various characteristic parameters reduce to their Newtonian limits.

Generally for $0 \leq k^2 \leq 1$, the incoming asymptote angle $\Psi_1$ can range from 0 to 90°, but the outgoing asymptote angle $\Psi_2$ given by eq.(26) can range from $2\pi - \Psi_1$ to $\infty$, because $\pi/2 \leq K(k) \leq \infty$ (for $0 \leq k^2 \leq 1$), $1/2\sqrt{2} = 0.353553 \leq \gamma \leq 1/2$ (for $0 \leq s \leq s_1$, and $1 \leq e \leq \infty$), and thus the particle coming along the incoming asymptote can go around the star or blackhole many times as its polar angle $\phi$ increases from $\Psi_1$ to $\Psi_2$ before going off to infinity along the outgoing asymptote. We have seen a similar behavior for a parabolic-type orbit [1] for which a particle coming from infinity at a polar angle $\phi = 0$ can go around the star or blackhole many times, before going off to infinity at a polar angle $\phi = 2K(k)/\gamma$.

For the special case of $k^2 = 1$ (where $s = s_1$) which is the boundary between Regions I and II, we find

$$e_1 = e_2 = \sqrt{\frac{g_2}{12}}, e_3 = -\sqrt{\frac{g_2}{3}}, \tag{35}$$

where $g_2 = 1/12 - s_1^2$ and $s_1^2$ is given by eq.(12), and the orbit equation becomes

$$\frac{1}{q} = \frac{1}{3} + 2\sqrt{\frac{g_2}{3}} \frac{1 - 5\sec h(2\gamma\phi)}{1 + \sec h(2\gamma\phi)}, \tag{36}$$

where $\gamma = (3g_2/4)^{1/4}$. The asymptote angle $\Psi_1$ is given by

$$\tanh^2(\gamma\Psi_1) = \frac{2}{3} - \frac{1}{18}\sqrt{\frac{3}{g_2}}. \tag{37}$$

From $g_2$ given by eq.(7) and the values of $s_1^2$ given at the beginning of Section III, we find that for $k^2 = 1$, $\Psi_1$ can range from 0 for $e = 1$ (parabolic-type orbit) to $2\tanh^{-1}(1/\sqrt{3}) = 1.31696 = 75.456°$ for $e = \infty$. Because $K(1) = \infty$, the orbit equation given by eq.(15) describes the trajectory of a particle that comes from infinity at an angle $\Psi_1$ to the horizontal axis and goes around the blackhole located at the origin counter-clockwise as $\phi$ increases, and finally circles around the blackhole with a radius that approaches $q_{\min}$ given by eq.(22). It is what we called an asymptotic hyperbolic orbit [1].

Thus the major characteristics of a hyperbolic-type orbit are described by the following set of parameters: its minimum distance $q_{\min}$ from the origin (where the star or blackhole is located) given by eq.(22), its two asymptote angles $\Psi_1$ and $\Psi_2$ relative to the horizontal axis given by eqs.(25) and (26), the intercepts $x_{a1}$ and $x_{a2}$ of these two asymptotes on the horizontal axis given by eq.(34), and its symmetry axis given by a straight line through the origin at an angle $\chi$ to the horizontal axis given by eq.(28). An asymptotic hyperbolic orbit having $k^2 = 1$ given by eq.(36) is described by one asymptote with the asymptote angle $\Psi_1$ given by eq.(37) that makes an intercept $x_{a1}$ on the horizontal axis given by eq.(34), and also by the radius of the spiraling circle around the blackhole that approaches $q_{\min}$ given by eq.(22) with $e_2$ given by eq.(35). The values of these characterizing parameters are presented in Tables II-VII for various values of $e$ from 1 to 10 and for various values of $k^2$ from 0 to 1. These tables should be used in conjunction with Table I that give the coordinates $(e, s)$ of the corresponding points. Note that the dimensionless distances $q_{\min}$, $x_{a1}$ and $x_{a2}$ are in units of the Schwarzschild radius $\alpha$ which depends on the mass $M$ of the star or blackhole corresponding to that particular coordinate point, and thus one should not compare these quantities at two different coordinate points just by their absolute values alone.

The Newtonian limit corresponds to the case of very small $s$ ($k^2 \simeq 0$) for which $sn(u, k) \simeq \sin u$, $\gamma \simeq 1/2$, $K(k) \simeq \pi/2$. The orbit equation (15) applies to the elliptic orbit for which $E_0 < 0$, $0 \leq e < 1$, the parabolic orbit for which $E_0 = 0$, $e = 1$, and the hyperbolic orbit for which $E_0 > 0$, $e > 1$. For the hyperbolic orbit, it follows from eqs.(25) that the equation for $\Psi_1$ becomes

$$\sin^2 \frac{\Psi_1}{2} = \frac{1}{2} - \frac{1}{2e}, \tag{38}$$

from which, denoting the angle $\Psi_1$ by $\phi_0$ in the Newtonian limit, we find

$$\phi_0 = \cos^{-1}\left(\frac{1}{e}\right), \tag{39}$$

in agreement with the well-known result from classical mechanics. The incoming asymptote makes an angle $\Psi_1 = \phi_0$, and the outgoing asymptote makes an angle $\Psi_2 = 2\pi - \phi_0$, with the horizontal axis, and the two asymptotes are symmetrical about the horizontal axis. The polar angle $\phi$ of the particle executing a Newtonian hyperbolic orbit is restricted to the range $\cos^{-1}(1/e) \leq \phi \leq 2\pi - \cos^{-1}(1/e)$ with $e > 1$. The minimum distance $r_{\min}$ of the particle from the star is given by

$$r_{\min} = \frac{h^2}{GM(1+e)}, \tag{40}$$

and it occurs at the polar angle $\phi = \pi$. The angle $\phi_0$ ranges from 0 for $e = 1$ to $\pi/2$ for $e \to \infty$.

The case for $k^2 = 1/2$ is also somewhat special in that it allows some characteristics of the orbit to be expressed more explicitly and simply. We give a brief discussion of $k^2 = 1/2$ in Appendix B.

To see the varied behavior of the hyperbolic-type orbit, we consider a given value of $e$ ($> 1$) and consider the orbits for increasing values of $s$ (and $k^2$) that starts from $s \simeq 0$ (and $k^2 \simeq 0$) to $s = s_1$ (for which $k^2 = 1$). The orbit for $s \simeq 0$ (and $k^2 \simeq 0$) is the Newtonian hyperbolic orbit with the asymptote angles $\Psi_1 \simeq \phi_0$, and $\Psi_2 = 2\pi - \phi_0$, giving the total deflection angle $\Theta \simeq 2\phi_0$. As $s$ increases, $k^2$ increases and $K(k)$ increases, and at some value of $s$ (and $k^2$), the outgoing asymptote becomes anti-parallel to the incoming asymptote. Beyond that value of $s$, the outgoing asymptote will start crossing the incoming asymptote as it goes to infinity, and as $s$ increases further, the particle can go around the star or blackhole an increasing number of times before it goes to infinity along the outgoing asymptote. Thus for the values of $s$ between 0 and $s_1$ for a given $e$, the particle can circle around the star or blackhole many times before going off to infinity, and there are infinitely many orbits for which the outgoing asymptotes can become parallel or anti-parallel to the incoming asymptotes. As $s$ becomes equal to $s_1$, the particle comes along the incoming asymptote and circles around the star with its radius approaching $q_{\min}$ given by eq.(22). In Figs.2-5, we present examples of these hyperbolic-type orbits, for a given value of $e = 2$, from a very small value of $s$ (for which $k^2 \simeq 0$) (Fig.2) coresponding to an orbit that is close to a Newtonian hyperbolic orbit, to $s = s_1 = 0.221035$ (for which $k^2 = 1$) (Fig.5) corresponding to the asymptotic hyperbolic-type orbit. In Fig.2, the orbit is for $k^2 = 0.001$, $s = 0.0111767$, and it can be checked from Tables II-IV that the symmetry axis makes an angle $180.07°$ with and thus almost coincides with the horizontal axis, and that the two asymptotes (dashed lines) make an angle $\Psi_1 = 59.994°$ and $\Psi_2 = 300.14°$ with the horizontal axis, i.e. close to $\phi_0 = 60°$ ($= \cos^{-1}(1/2)$) and $300°$ ($= 360° - \phi_0$) respectively for the corresponding asymptote angles for Newtonian hyperbolic orbit, and that the three axes are concurrent at an intersection point close to the horizontal axis. Tables VI and VII give the $x$-intercept of the first (incoming) asymptote to be $x_{a1} = -2668.6$ and that of the second (outgoing) asymptote to be $x_{a2} = -2672.2$, and Table V gives the minimum distance of the hyperbolic-type orbit from the origin to be $q_{\min} = 1333.5$ (all in units of Schwarzschild radii), while the corresponding Newtonian hyperbola for the same values of $e$ and $s$ would give values of $q_{\min} = 1334.2$ and $x_{a1} = x_{a2} = -2668.4$ (from the formulas given in Appendix A). The orbit in Fig.3 for $k^2 = 0.3$ differs more from the Newtonian hyperbolic orbit, and the concurrent point of the two asymptotes and the symmetry axis (dashed lines) is away from the horizontal axis. The orbit in Fig.4 for $k^2 = 0.7$ shows that the trajectory intersects itself. The two asymptotes and the symmetry axis are always concurrent except for the parallel and anti-parallel case discussed above. The intercept of the outgoing asymptote is a positive value ($x_{a2} = 8.3133$ from Table VII) here. Fig.5 shows an asymptotic hyperbolic-type orbit for which the particle comes from infinity at an angle of $58.04°$ to the horizontal axis and circles the star or blackhole counter-clockwise with a radius that approaches $q_{\min} = 1.8257$.

We note that as $e$ increases, the range of $s$ in Region I which gives these hyperbolic-type orbits between 0 and $s_1$ becomes smaller and smaller. However, as $s$ increases from 0 to $s_1$ for a fixed value of $e$, the squared modulus $k^2$ of the elliptic functions that describe the orbit still changes from 0 to 1, and the corresponding orbit still undergoes changes similar to the sequence given in Fig.2-5 for $e = 2$.

One is often tempted to compare these hyperbolic-type trajectories of a nonzero mass particle with the trajectories of light in the gravitational field. Exact expressions for the trajectories of light in the Schwarzschild metric have been presented in ref.1. We should note that while the trajectories of light can be characterized by a single parameter which can be taken to be the distance of closest approach between the light path and the center of the star or blackhole, the trajectories of a particle with a nonzero mass are characterized by two parameters $e$ and $s$, i.e. different sets of $(e, s)$ corresponding to different hyperbolic-type trajectories can have the same $q_{\min}$, the distance of closest approach of the particle from the star or blackhole. We can discuss and compare the small deviations due to general relativity for light from its straight line path, and for a nonzero mass particle from its Newtonian hyperbolic path. We shall make such comparisons in the following section.

## 4 Small Deviations from Newtonian Hyperbolic Orbit

For the case of very small $k^2$ and $s$ when the general relativistic hyperbolic-type orbit deviates only slightly from the Newtonian hyperbolic orbit, besides comparing the angle $\Psi_1$ with $\phi_0$, there are other angles of interest for comparison and we show them in Fig.6 where the thick solid curve represents the general relativistic hyperbolic-type orbit and the thick dashed curve represents the corresponding Newtonian hyperbolic orbit. Besides $\Psi_1$ and $\Psi_2$ given by eqs.(25) and (26) and $\phi_0$ given by eq.(39), an angle that is complementary to $\Psi_1$ is

$$\Psi_1' \equiv 2\pi - \Psi_2 = 2\pi - \left(\frac{2K(k)}{\gamma} - \Psi_1\right). \tag{41}$$

The other useful angles are defined by

$$\delta \equiv \phi_0 - \Psi_1, \tag{42}$$

$$\delta' \equiv \phi_0 - \Psi_1' = \delta + \frac{2K(k)}{\gamma} - 2\pi, \tag{43}$$

$$\Theta_{GR} \equiv \Psi_1 + \Psi_1', \tag{44}$$

$$\Theta_{Newton} \equiv 2\phi_0, \tag{45}$$

and

$$\Delta\phi \equiv \Theta_{Newton} - \Theta_{GR} = \delta + \delta'$$
$$= \frac{2K(k)}{\gamma} - 2\pi + 2\phi_0 - 2\Psi_1, \tag{46}$$

where the subscript $Newton$ for the Newtonian limit has been used in contrast to the general relativistic case with the subscript $GR$. The Newtonian hyperbolic orbit is symmetric about the horizontal axis but the general relativistic hyperbolic-type orbit is generally not symmetric about the horizontal axis, but is symmetric about a line making an angle $K(k)/\gamma$ with the horizontal axis, and hence there are differences between $\delta$ and $\delta'$ and between $\Psi_1$ and $\Psi'_1$. We shall now derive the approximate expressions for the above angles.

While the special case of $k^2 = 0$ (which implies $s = 0$ and $q = \infty$), is not interesting because it is the limiting case of a particle in a zero gravitational field, the case of $k^2 \simeq 0$ and $s \simeq 0$ for which $q$ is large but finite, is very important because it covers the situation in which there are small deviations from Newtonian mechanics due to general relativity. We now consider the small deviations from the Newtonian hyperbolic orbit.

We begin by expanding $\cos\theta$ of eq.(20) in power series in $s$, using $g_2$ and $g_3$ given by eq.(7), and find

$$\cos\theta = 1 - 2 \cdot 3^3 e^2 s^4 - 2^2 \cdot 3^3 (1 + 9e^2) s^6 + .. \tag{47}$$

from which we find

$$\theta = 2\sqrt{27} e s^2 [1 + (9 + e^{-2}) s^2 + ..]. \tag{48}$$

From eq.(19), we then find [8]

$$\begin{aligned}
e_1 &= \frac{1}{6} - s^2 - (e^2 + 3)s^4 + .. \\
e_2 &= -\frac{1}{12} + \frac{1}{2}(e+1)s^2 + \frac{(e+1)^3}{2e} s^4 + .. \\
e_3 &= -\frac{1}{12} - \frac{1}{2}(e-1)s^2 + \frac{(e-1)^3}{2e} s^4 + ...
\end{aligned} \tag{49}$$

From eq.(49), the right hand side of eq.(25) can be approximated by

$$\sqrt{-\frac{\frac{1}{3} + 4e_3}{4(e_2 - e_3)}} = \sqrt{\frac{e-1}{2e}} \left\{ 1 - \frac{(e+1)(e^2+1)}{2e^2} s^2 + ... \right\}. \tag{50}$$

We notice that both expansions of the numerator and denominator on the left-hand side of eq.(50) start with terms in $s^2$, and hence we need $e_2$ and $e_3$ expanded to the powers of $s^4$ from eq.(19) in order to get the right-hand side of eq.(50) to be correctly expanded to the power of $s^2$. This requires $\theta$ to be expanded to the power of $s^4$, and to obtain the coefficient of $s^4$ in the expansion of $\theta$, we equate $\cos\theta = 1 - \theta^2/2 + ..$ and find that we require an expansion of $\cos\theta$ to the power of $s^6$ that is presented in eq.(47).

To approximate the left hand side of eq.(25), we use the Fourier expansion of the Jacobian elliptic function $sn(u,k)$ given by [4]

$$sn(u,k) = \frac{2\pi}{kK(k)} \left\{ \frac{q^{1/2}}{1-q} \sin\frac{\pi u}{2K(k)} + \frac{q^{3/2}}{1-q^3} \sin\frac{3\pi u}{2K(k)} + ... \right\}, \tag{51}$$

where

$$q = \frac{k^2}{16}\left[1 + 8\left(\frac{k}{4}\right)^2 + ...\right]. \tag{52}$$

From eqs.(16), (17) and (49), $k^2$ and $\gamma$ are given by

$$k^2 = 4es^2\left(1 + \frac{1 + 9e^2 - 2e^3}{e^2}s^2 + ..\right), \tag{53}$$

$$\gamma = \frac{1}{2}[1 - (3 - e)s^2 + ..], \tag{54}$$

and from [4]

$$K(k) = \frac{\pi}{2}\left(1 + \frac{k^2}{4} + ...\right) \tag{55}$$

and eq.(17), we find

$$K(k) = \frac{\pi}{2}(1 + es^2 + ..), \tag{56}$$

$$\frac{2K(k)}{\gamma} = 2\pi(1 + 3s^2 + ..). \tag{57}$$

Using eqs.(51) to (57), and writing $\Psi_1 = \phi_0 - \delta$ as in eq.(25), where $\delta$ is small, the left hand side of eq.(25) can be approximated, to the order of $\delta$ and $s^2$, by

$$sn(\Psi_1, k) = \sin\frac{1}{2}\phi_0 - \frac{1}{2}\delta\cos\frac{1}{2}\phi_0 + \frac{1}{4}es^2\left(\sin\frac{1}{2}\phi_0 + \sin\frac{3}{2}\phi_0\right) - \frac{3}{2}s^2\phi_0\cos\frac{1}{2}\phi_0 + ... \tag{58}$$

Equating eqs.(58) and (50), and from eq.(43), we find

$$\delta = [(2 + e^{-2})\sqrt{e^2 - 1} - 3\phi_0]s^2 + .., \tag{59}$$

$$\delta' = \delta + 6\pi s^2 + .., \tag{60}$$

and from eqs.(42), (41) and (57), we find

$$\Psi_1 = \phi_0 - [(2 + e^{-2})\sqrt{e^2 - 1} - 3\phi_0]s^2 + .., \tag{61}$$

$$\Psi_1' = \Psi_1 - 6\pi s^2 + .., \tag{62}$$

From eqs.(44) and (45), we finally arrive at

$$\Theta_{GR} = 2\phi_0 - [6\pi - 6\phi_0 + 2(2 + e^{-2})\sqrt{e^2 - 1}]s^2 + ..., \tag{63}$$

and

$$\Delta\phi \equiv \Theta_{Newton} - \Theta_{GR} = [6\pi - 6\phi_0 + 2(2 + e^{-2})\sqrt{e^2 - 1}]s^2 + ... \quad (64)$$

for $s \ll s_1$, where $s_1$ is given by eq.(12). Equation (64) is the main result of this section. It represents the deviation from the Newtonian hyperbolic trajectory according to general relativity in the Schwarzschild metric for the case of a weak gravitational field that can perhaps be experimentally checked for objects in our solar system. For a parabolic-type orbit, $e = 1$ and $\phi_0 = 0$, and the deviation $\Delta\phi \simeq 6\pi s^2$ from a Newtonian parabolic orbit is the same as the precession angle of a general relativistic elliptic-type orbit. The first term $6\pi s^2$ of eq.(64) comes from the precession angle given by eq.(21). For a general relativistic hyperbolic-type orbit (for which $e > 1$ and $\phi_0 > 0$), the deviation $\Delta\phi$ from the Newtonian hyperbolic trajectory involves a couple of additional terms that depend on $e$, as shown in eq.(64).

It is interesting to compare this deviation with the deflection of light according to general relativity. The deflection of light due to gravity according to general relativity is considered as a deviation from a straight line path, and the deviation $\Delta\phi$ above is the deviation of the path of a particle (with a nonzero mass) according to general relativity from the Newtonian hyperbolic path, not a straight line path. According to general relativity, light not only can be deflected, but also can be significantly deflected, turned around, and/or captured in the presence of a gravitational field. Here we want to compare only the case involving small deviations. The deviation for a particle with a nonzero mass is given by eq.(64), and the deviation of light from a straight line path is well known to be

$$\Delta\phi_{light} \simeq \frac{4GM}{c^2 R} = \frac{2\alpha}{R}, \quad (65)$$

where $R$ is the distance of closest approach of the path of light from the center of the star or blackhole, and $\alpha$ is given by eq.(3). To make a comparison of eqs.(64) and (65), we need to equate $R$ with $r_{\min}$, the distance of closest approach the particle making the hyperbolic-type trajectory about the star. From eqs.(22) and (49), we have

$$\frac{1}{q_{\min}} = \frac{\alpha}{r_{\min}} = 2(e + 1)s^2 + ... \quad (66)$$

Setting $R = r_{\min}$ in eq.(65) and using eq.(66), we can express $\Delta\phi_{light}$ of eq.(65) as

$$\Delta\phi_{light} = 4(e + 1)s^2 + ... \quad (67)$$

that would allow us to compare the deviation of light from a straight line trajectory with the deviation of a nonzero mass particle from its Newtonian hyperbolic trajectory corresponding to some values of $e$ and $s$ for which the particle has the same distance of closest approach given by eq.(66) from the star or blackhole. Equations (67) and (64) give such a comparison of the deviations of light and a nonzero mass particle from their straight line and Newtonian hyperbolic trajectories respectively. It should be remembered that $s$ should be much less than $s_1$ given by eq.(12). In particular, for the case of very large $e$ so that the particle's hyperbolic orbit is close to a straight line with $\phi_0 \simeq \pi/2$, then we have, from eqs.(64) and (67),

$$\Delta\phi \simeq (6\pi - 6\phi_0 + 4e)s^2, \tag{68}$$

compared to

$$\Delta\phi_{light} \simeq 4es^2. \tag{69}$$

$\Delta\phi$ approximated by eq.(68) is the deviation of the trajectory of a particle with a nonzero mass according to general relativity from the Newtonian hyperbolic trajectory that is close to but not a straight line trajectory, while $\Delta\phi_{light}$ approximated by eq.(69) is the deviation of the trajectory of light according to general relativity from a straight line trajectory. The requirement that $s << s_1$ or $es^2 << (\sqrt{27})^{-1}$ in eqs.(68) and (69) should be remembered.

We note that this comparison differs from one that is often used and quoted in many texts [9] in which the deflection of light from its straight line trajectory given by eq.(65) according to general relativity is compared with the "corresponding deflection" $\theta$ of a particle with a mass from its straight line trajectory by a gravitational field according to Newtonian mechanics, where $\theta$ is the analog of the Rutherford scattering angle for an attractive potential and is given, for small $\theta$, by

$$\theta \simeq \frac{2GM}{v_0^2 b}, \tag{70}$$

where $v_0^2/2$ is the initial kinetic energy per unit mass of the particle and $b$ is the impact parameter. As $b$ is related to the distance of closest approach $R$ by

$$b = R\sqrt{1 + \frac{2GM}{v_0^2 R}}, \tag{71}$$

we find, if $v_0$ can be replaced by $c$, the speed of light, that the deflection angle $\theta$ of light would be given by

$$\theta \simeq \frac{2GM}{c^2 R}, \tag{72}$$

and thus Newtonian mechanics would have predicted a bending of light that would be only half of the experimentally observed angle $\Delta\phi$ given by eq.(65) according to general relativity. However, replacing $v_0$ by $c$ while keeping the non-relativistic expression for the kinetic energy of the particle is not consistent with a fully general relativistic approach.

We make note of eqs.(70)-(72) and mention the comparison of eq.(72) with eq.(65) here, which is often quoted in many texts and is well known, so that the comparison between eqs.(67) and (64), and between eqs.(69) and (68), that we have made can be more fully appreciated.

We now apply our approximation formula given by eq.(64) to a possible experimental test of general relativity that can be carried out in our solar system. Longuski et al [2] presented an approximate formula which they used to give a general relativistic deflection of a spacecraft as a test of general relativity as well as a possible check on the two constants that arise in the parametrized post-Newtonian metric. The proposed path of the spacecraft involves a close pass of the Sun or the planet Jupiter or the Earth. We have used the same input data for the initial speeds $v_\infty$ of the spacecraft (when it is very far away from the planet or star) and the minimum distances of approach $r_p$ (that we call $r_{\min}$ in our previous description) of the spacecraft from the respective planet or star that Longuski et al used, but instead apply our approximate formula given by eq.(64) for calculating the deviations $\Delta\phi$ of the trajectories of the spacecraft from their Newtonian paths. Our calculations are carried out in the following steps. We first calculate $v_p$, the velocity of the spacecraft when it is at the distance of the closest approach from the star or planet, from conservation of total energy, that is given by

$$\frac{v_p^2}{c^2} = \frac{v_\infty^2}{c^2} + \left(\frac{2GM}{c^2}\right)\frac{1}{r_p} - \left(\frac{GM}{c^2}\right)^2 \frac{1}{r_p^2} + ..., \tag{73}$$

where $M$ is the mass of the star or planet. Then from $h = r_p v_p / \sqrt{1 - v_p^2/c^2}$, the angular momentum of the spacecraft per unit rest mass of the spacecraft, we find, from eq.(9), that

$$s = \frac{GM}{c^2} \frac{1}{(v_p/c)r_p} \sqrt{1 - \frac{v_p^2}{c^2}}. \tag{74}$$

We proceed to calculate $e$ according to eq.(31). Some of the terms presented above do not give any significant contributions for the input data for this particular example, but we have presented the above expressions for possible future use. The Newtonian asymptote angle $\phi_0$ is calculated from eq.(39). We then use the calculated values of $e$, $s$, and $\phi_0$, and eq.(64) to obtain $\Delta\phi$, and present our results in Table VIII, where the first three rows are the same input data presented in Longuski et al [2].

Our values for $\Delta\phi$ can be compared with the values of $\Delta\phi_E$ given by Longuski et al (where the subscript $E$ stands for Einstein to represent the case in which both post-Newtonian parameters are set equal to one in their formula) that are equal to $3.229 \times 10^{-9}$, $1.767 \times 10^{-7}$ and $4.673 \times 10^{-6}$ for the Earth, Jupiter and Sun respectively. It is seen that our values are somewhat larger than theirs.

# 5 Terminating Orbit for e>1

The terminating orbit for $e > 1$ in Region I is given by the same expression as that for $0 \leq e \leq 1$ and is given by [1]

$$\frac{1}{q} = \frac{1}{3} + 4\frac{e_1 - e_2 sn^2(\gamma\phi, k)}{cn^2(\gamma\phi, k)}, \tag{75}$$

where $\gamma, k, e_1, e_2, e_3$ are given by eqs.(17)-(19). Unlike the case for $0 \leq e \leq 1$ for which the particle starts from the polar angle $\phi = 0$ in a direction perpendicular to the line joining it to the star or blackhole, for the case $e > 1$, the particle starts from infinity at a polar angle $\phi = \Psi_1$, where $\Psi_1$ is given by eq.(25). The particle plunges into the center of the blackhole when its polar angle reaches $\phi_1$, where $\phi_1$ is given by

$$\phi_1 = \frac{K(k)}{\gamma}. \tag{76}$$

Examples of these terminating orbits for $0 \leq e \leq 1$ were given in ref.1, and they are similar for $e > 1$.

We shall discuss the terminating orbit in Region II (for $s > s_1$) for $e > 1$ in another paper for which the range of $k^2$ is different from that for $0 \leq e \leq 1$.

# 6 Summary

We have presented exact analytic results for various characteristics of the hyperbolic-type orbits of a particle in the gravitational field produced by a planet, star or blackhole, in the Schwarzschild metric, among them the characteristics of the asymptotes given by eqs.(25), (26) and (34), and of the symmetry axis given by eq.(28). Tables and graphs are included which clearly show how these characteristics vary as the energy and field parameters $e$ and $s$ vary. We have also derived a simple approximation formula, eq.(64), for the case when the deviation from the classical Newtonian path is small, which can have useful applications for experiments involving tests of general relativity that are carried out in our solar system.

# 7 Appendix A The Asymptotes and their Intercepts in the Newtonian Limit

In this Appendix, we briefly describe the Newtonian limit of the hyperbolic-type orbit given by eq.(15). This is done by setting $s^2 \simeq 0$ and $k^2 \simeq 0$ and using all the approximate formulas given in eqs.(53)-(57), and (49), and the relations given by eqs.(17), (18), (13) and noting $sn(x, k) \simeq \sin x$, $cn(s, k) \simeq \cos x$, and $dn(x, k) \simeq 1$, for small $k^2$. We readily arrive at eq.(24) from eq.(15) in the Newtonian limit. We now consider how to relate eq.(24) to the classical theory of conic sections that gives the equations for the ellipse, parabola and hyperbola in polar coordinates as

$$\frac{1}{r} = \frac{1}{A}(1 - e\cos\phi), \tag{77}$$

where $A = a(1 - e^2)$ for an ellipse [for which the other two relevant parameters are $b = a\sqrt{1 - e^2}$ and $c = ae = \sqrt{a^2 - b^2}$ ($c$ is not the speed of light here)], and $A = a(e^2 - 1)$ for a hyperbola (for which the other two relevant parameters are $b = a\sqrt{e^2 - 1}$ and $c = ae = \sqrt{a^2 + b^2}$). The parameters $a, b, c$ have specific meanings when the curves are described in Cartesian coordinates. Let $C$ be the center of the Cartesian coordinate system used to describe these curves. Then the origin $O$ of our polar coordinate system where the star or blackhole is located is a focus of the curve. The distance $CO$ is equal to $ae$ and the minimum distance of the curve from $O$ is equal to $a(1 - e)$ for an ellipse, $A/2$ for a parabola, and $a(e - 1)$ for a hyperbola. The two asymptotes of a hyperbola makes an angle $\phi_0$ and $2\pi - \phi_0$ with the horizontal axis, where

$$\phi_0 = \cos^{-1}\left(\frac{1}{e}\right) = \cos^{-1}\left(\frac{a}{c}\right). \tag{78}$$

Equating eq.(77) to eq.(24), we find

$$a = \frac{h^2}{GM\,|\,1 - e^2\,|} = \frac{GMm}{2\,|\,E_0\,|}, \tag{79}$$

or

$$\frac{a}{\alpha} = \frac{1}{|\,2s^2(1 - e^2)\,|} \tag{80}$$

for an ellipse or a hyperbola, and

$$A = \frac{h^2}{GM} \tag{81}$$

or

$$\frac{A}{\alpha} = \frac{1}{2s^2} \tag{82}$$

for a parabola.

We now consider the hyperbolic orbit only. The impact parameter $b$ is the perpendicular distance from $O$ to either one of the asymptotes, and it is given by

$$b = ae\sin\phi_0 = \frac{h^2}{GM}\frac{1}{\sqrt{e^2 - 1}}. \tag{83}$$

Notice from eq.(83) and (79) that $b = a\sqrt{e^2 - 1}$, i.e. the impact parameter $b$ is equal to the parameter $b$, one of the three parameters $a, b, c$ that are used for characterizing a hyperbolic curve, a point that may not have been often noted. From eq.(83), we find that the dimensionless impact parameter is given by

$$\frac{b}{\alpha} = \frac{1}{2s^2\sqrt{e^2-1}}, \qquad (84)$$

which, surprisingly, agrees exactly with the expression for the dimensionless impact parameter given by eq.(33) which we derived earlier using relativity. The horizontal axis is the symmetry axis of the hyperbolic curve, and it and the two asymptotes are concurrent at a point that we called $C$, the center of the Cartesian coordinate system that we use to describe the hyperbolic curve. The distance of $C$ from $O$ is the distance of the intersection point of the two asymptotes on the horizontal axis from the origin and is given by

$$r_{a1} = r_{a2} = ae = c = \frac{h^2}{GM}\frac{e}{e^2-1}, \qquad (85)$$

and thus the dimensionless $x$ coordinates of the intersection points of the two asymptotes on the horizontal axis are given by

$$x_{a1} = x_{a2} = -\frac{e}{2s^2(e^2-1)}. \qquad (86)$$

The minimum distance $r_{\min}$ of the hyperbolic curve from the origin $O$ is

$$r_{\min} = a(e-1) = \frac{h^2}{GM(e+1)}, \qquad (87)$$

and hence

$$q_{\min} = \frac{1}{2s^2(e+1)}. \qquad (88)$$

We can now relate the asymptote angles, the symmetry axis, the $x$ intercepts of the asymptotes, and the minimum distance of the hyperbolic-type orbit discussed in the text, to their corresponding Newtonian limits. For $s \simeq 0$ and $k^2 \simeq 0$, it follows from eqs.(25), (26), (50) and (57), that we have $\Psi_1 \simeq \phi_0$, $\Psi_2 \simeq 2\pi - \phi_0$, where $\phi_0$ is given by eq.(78). The symmetry axis of the hyperbolic-type orbit that makes an angle $\chi$ given by eq.(28) becomes $\chi \simeq \pi$ and thus it coincides with the horizontal axis. The $x$ intercepts of the two asymptotes from eq.(34) become $x_{a1} \simeq x_{a2}$ and are given by eq.(86) (see Fig.2). Using eq.(49), $q_{\min}$ given by eq.(22) is approximated by eq.(88).

# 8  Appendix B The Special Case of $k^2 = 1/2$

In this Appendix, we show that for the case of $k^2 = 1/2$ the asymptote angles given by eqs.(25) and (26) can be expressed more explicitly.

It was shown in ref.1 that for $k^2 = 1/2$ in Region I, we have $e_1 = -e_3 = \sqrt{g_2}$, $e_2 = 0$, $\gamma = \sqrt[4]{g_2}$, where $g_2$ is given by eq.(7), and $s$ and $e$ are related by

$$s^2 = \frac{1}{6(e^2-1)}\left(\sqrt{\frac{1+2e^2}{3}}-1\right). \tag{89}$$

Thus the orbit equation (15) becomes

$$\frac{1}{q} = \frac{1}{3} - 2\sqrt{g_2}cn^2(\gamma\phi, 1/\sqrt{2}), \tag{90}$$

with $q_{\min} = 3$. The asymptote angles become

$$\Psi_1 = \frac{1}{\sqrt[4]{g_2}}sn^{-1}\left(\sqrt{1-\frac{1}{6\sqrt{g_2}}}, \frac{1}{\sqrt{2}}\right), \tag{91}$$

and

$$\Psi_2 = \frac{2K(1/\sqrt{2})}{\sqrt[4]{g_2}} - \Psi_1, \tag{92}$$

where $K(1/\sqrt{2}) = 1.8540747$, and the symmetry axis through the origin makes an angle $\chi$ with the horizontal axis given by

$$\chi = \frac{K(1/\sqrt{2})}{\sqrt[4]{g_2}}. \tag{93}$$

The terminating orbit equation in Region I for $k^2 = 1/2$ becomes

$$\frac{1}{q} = \frac{2\sqrt{g_2}}{cn^2(\gamma\phi, 1/\sqrt{2})}, \tag{94}$$

with $q_1$ and $\phi_1$ given by

$$\frac{1}{q_1} = \frac{1}{3} + 2\sqrt{g_2}, \tag{95}$$

and

$$\phi_1 = \frac{K(1/\sqrt{2})}{\sqrt[4]{g_2}}. \tag{96}$$

# 9 References

*Electronic address: fhioe@sjfc.edu

[1] F.T. Hioe and D. Kuebel, Phys. Rev. D 81, 084017 (2010).

[2] J.M. Longuski, E. Fischbach, and D.J. Scheeres, Phys. Rev. Lett. 86, 2942 (2001), J.M. Longuski, E. Fischbach, D.J. Scheeres, G. Giampierri, and R.S. Park, Phys. Rev. D 69, 042001 (2004).

[3] M.P. Hobson, G. Efstathiou and A.N. Lasenby: General Relativity, Cambridge University Press, 2006, Chapters 9 and 10.

[4] P.F. Byrd and M.D. Friedman: Handbook of Elliptic Integrals for Engineers and Scientists, 2nd Edition, Springer-Verlag, New York, 1971.

[5] Equations similar to eqs.(15) and (16) in somewhat different forms were given by A.R. Forsyth, Proc. Roy. Soc. Lond. A97, 145 (1920), F. Morley, Amer. J. Math. 43, 29 (1921), E.T. Whittaker: A Treatise on the Analytical Dynamics of Particles and Rigid Bodies, 4th Edition, Dover, New York, 1944, Chapter XV, and C.G. Darwin, Proc. Roy. Soc. Lond. A249, 180 (1958), ibid. A263, 39 (1961). Their analyses of these equations were more restricted than ours, and except for a short section in Darwin, they did not discuss the hyperbolic-type orbit. See also a brief section on the unbound orbits in S. Chandrasekhar: The Mathematical Theory of Black Holes, Oxford University Press, 1992, Section 19 (c).

[6] F.T. Hioe and D. Kuebel, in preparation.

[7] The misprint $s_1 \to (\sqrt{27}e)^{-1}$ on the 5th line of Section V in ref.1 should be corrected to $s_1^2 \to (\sqrt{27}e)^{-1}$.

[8] The coefficients of $s^4$ in $e_2$ and $e_3$ in eq.(20) in F.T. Hioe, Phys. Lett. A 373, 1506 (2009) are incorrect and should be replaced by those given in eq.(49) of this paper. However these errors did not affect any other results presented in that earlier paper.

[9] See e.g. J. Plebanski and A. Krasinski: An Introduction to General Relativity and Cosmology, Cambridge University Press, 2006, p. 185.

Table I. Values of s for various values of $e$ and $k^2$ in Region I.

| $s$ | $e = 1$ | $e = 2$ | $e = 3$ | $e = 4$ | $e = 5$ | $e = 6$ | $e = 7$ | $e = 8$ | $e = 9$ | $e = 10$ |
|---|---|---|---|---|---|---|---|---|---|---|
| $k^2 = 0.001$ | 0.0157956 | 0.0111767 | 0.00912752 | 0.00790543 | 0.00707124 | 0.00645537 | 0.00597667 | 0.00559078 | 0.00527112 | 0.00500069 |
| $k^2 = 0.01$ | 0.0495050 | 0.0352386 | 0.0288293 | 0.0249910 | 0.0223654 | 0.0204245 | 0.0189145 | 0.0176965 | 0.0166871 | 0.0158327 |
| $k^2 = 0.1$ | 0.143740 | 0.107928 | 0.0898423 | 0.0785577 | 0.0706709 | 0.0647625 | 0.0601240 | 0.0563574 | 0.0532200 | 0.0505542 |
| $k^2 = 0.2$ | 0.186339 | 0.146669 | 0.124286 | 0.109692 | 0.0992460 | 0.0913021 | 0.0850004 | 0.0798444 | 0.0755247 | 0.0718374 |
| $k^2 = 0.3$ | 0.210663 | 0.171942 | 0.148008 | 0.131757 | 0.119859 | 0.110677 | 0.103319 | 0.0972536 | 0.0921424 | 0.0877595 |
| $k^2 = 0.4$ | 0.225877 | 0.189455 | 0.165246 | 0.148218 | 0.135493 | 0.125540 | 0.117488 | 0.110804 | 0.105141 | 0.100263 |
| $k^2 = 0.5$ | 0.235702 | 0.201667 | 0.177753 | 0.160438 | 0.147269 | 0.136850 | 0.128352 | 0.121253 | 0.115209 | 0.109984 |
| $k^2 = 0.6$ | 0.242061 | 0.210023 | 0.186583 | 0.169227 | 0.155844 | 0.145156 | 0.136380 | 0.129012 | 0.122715 | 0.117254 |
| $k^2 = 0.7$ | 0.246076 | 0.215507 | 0.192512 | 0.175212 | 0.161738 | 0.150904 | 0.141963 | 0.134429 | 0.127971 | 0.122357 |
| $k^2 = 0.8$ | 0.248452 | 0.218833 | 0.196166 | 0.178937 | 0.165430 | 0.154521 | 0.145489 | 0.137859 | 0.131306 | 0.125600 |
| $k^2 = 0.9$ | 0.249653 | 0.220540 | 0.198058 | 0.180877 | 0.167361 | 0.156418 | 0.147342 | 0.139665 | 0.133064 | 0.127312 |
| $k^2 = 1.0$ | 0.250000 | 0.221035 | 0.198609 | 0.181444 | 0.167926 | 0.156974 | 0.147886 | 0.140195 | 0.133581 | 0.127815 |

Table II. Values of $\chi$ (see eq.28) in radians for various values of $e$ and $k^2$ in Region I.

| $\chi$ | $e = 1$ | $e = 2$ | $e = 3$ | $e = 4$ | $e = 5$ | $e = 6$ | $e = 7$ | $e = 8$ | $e = 9$ | $e = 10$ |
|---|---|---|---|---|---|---|---|---|---|---|
| $k^2 = 0.001$ | 3.1439 | 3.1428 | 3.1424 | 3.1422 | 3.1421 | 3.1420 | 3.1419 | 3.1419 | 3.1419 | 3.1418 |
| $k^2 = 0.01$ | 3.1652 | 3.1535 | 3.1495 | 3.1476 | 3.1464 | 3.1456 | 3.1450 | 3.1446 | 3.1443 | 3.1440 |
| $k^2 = 0.1$ | 3.3823 | 3.2706 | 3.2310 | 3.2109 | 3.1988 | 3.1907 | 3.1849 | 3.1805 | 3.1771 | 3.1744 |
| $k^2 = 0.2$ | 3.6361 | 3.4240 | 3.3448 | 3.3040 | 3.2792 | 3.2625 | 3.2505 | 3.2415 | 3.2345 | 3.2289 |
| $k^2 = 0.3$ | 3.9083 | 3.6057 | 3.4877 | 3.4260 | 3.3882 | 3.3626 | 3.3443 | 3.3304 | 3.3196 | 3.3109 |
| $k^2 = 0.4$ | 4.2064 | 3.8213 | 3.6657 | 3.5832 | 3.5324 | 3.4979 | 3.4730 | 3.4542 | 3.4395 | 3.4277 |
| $k^2 = 0.5$ | 4.5415 | 4.0795 | 3.8876 | 3.7847 | 3.7209 | 3.6776 | 3.6462 | 3.6224 | 3.6038 | 3.5888 |
| $k^2 = 0.6$ | 4.9321 | 4.3953 | 4.1677 | 4.0447 | 3.9681 | 3.9158 | 3.8779 | 3.8491 | 3.8266 | 3.8084 |
| $k^2 = 0.7$ | 5.4119 | 4.7974 | 4.5332 | 4.3895 | 4.2997 | 4.2383 | 4.1936 | 4.1598 | 4.1332 | 4.1118 |
| $k^2 = 0.8$ | 6.0567 | 5.3516 | 5.0458 | 4.8788 | 4.7741 | 4.7025 | 4.6504 | 4.6108 | 4.5797 | 4.5547 |
| $k^2 = 0.9$ | 7.1073 | 6.2694 | 5.7045 | 5.5791 | 5.3158 | 5.4932 | 5.4307 | 5.3832 | 5.3459 | 5.3158 |
| $k^2 = 1.0$ | — | — | — | — | — | — | — | — | — | — |

Table III. Values of $\Psi_1$ (see eq.25) in radians for various values of $e$ and $k^2$ in Region I.

| $\Psi_1$ | $e = 1$ | $e = 2$ | $e = 3$ | $e = 4$ | $e = 5$ | $e = 6$ | $e = 7$ | $e = 8$ | $e = 9$ | $e = 10$ |
|---|---|---|---|---|---|---|---|---|---|---|
| $k^2 = 0.001$ | 0 | 1.0471 | 1.2308 | 1.3179 | 1.3691 | 1.4030 | 1.4271 | 1.4451 | 1.4591 | 1.4702 |
| $k^2 = 0.01$ | 0 | 1.0463 | 1.2291 | 1.3156 | 1.3665 | 1.4001 | 1.4240 | 1.4418 | 1.4557 | 1.4667 |
| $k^2 = 0.1$ | 0 | 1.0386 | 1.2132 | 1.2943 | 1.3414 | 1.3723 | 1.3942 | 1.4105 | 1.4231 | 1.4331 |
| $k^2 = 0.2$ | 0 | 1.0315 | 1.1979 | 1.2733 | 1.3165 | 1.3446 | 1.3643 | 1.3789 | 1.3901 | 1.3990 |
| $k^2 = 0.3$ | 0 | 1.0259 | 1.1853 | 1.2556 | 1.2952 | 1.3206 | 1.3382 | 1.3512 | 1.3612 | 1.3691 |
| $k^2 = 0.4$ | 0 | 1.0216 | 1.1751 | 1.2411 | 1.2776 | 1.3006 | 1.3165 | 1.3280 | 1.3368 | 1.3438 |
| $k^2 = 0.5$ | 0 | 1.0184 | 1.1674 | 1.2298 | 1.2637 | 1.2848 | 1.2991 | 1.3095 | 1.3173 | 1.3234 |
| $k^2 = 0.6$ | 0 | 1.0161 | 1.1617 | 1.2215 | 1.2533 | 1.2728 | 1.2860 | 1.2954 | 1.3024 | 1.3079 |
| $k^2 = 0.7$ | 0 | 1.0146 | 1.1578 | 1.2157 | 1.2461 | 1.2687 | 1.2768 | 1.2855 | 1.2920 | 1.2969 |
| $k^2 = 0.8$ | 0 | 1.0137 | 1.1554 | 1.2120 | 1.2415 | 1.2592 | 1.2709 | 1.2792 | 1.2853 | 1.2900 |
| $k^2 = 0.9$ | 0 | 1.0132 | 1.1541 | 1.2101 | 1.2391 | 1.2564 | 1.2678 | 1.2758 | 1.2817 | 1.2863 |
| $k^2 = 1.0$ | 0 | 1.0130 | 1.1538 | 1.2096 | 1.2384 | 1.2556 | 1.2669 | 1.2749 | 1.2807 | 1.2852 |

Table IV. Values of $\Psi_2$ (see eq.26) in radians for various values of $e$ and $k^2$ in Region I.

| $\Psi_2$ | $e = 1$ | $e = 2$ | $e = 3$ | $e = 4$ | $e = 5$ | $e = 6$ | $e = 7$ | $e = 8$ | $e = 9$ | $e = 10$ |
|---|---|---|---|---|---|---|---|---|---|---|
| $k^2 = 0.001$ | 6.2879 | 5.2384 | 5.0540 | 4.9665 | 4.9150 | 4.8809 | 4.8568 | 4.8387 | 4.8246 | 4.8134 |
| $k^2 = 0.01$ | 6.3304 | 5.2607 | 5.0700 | 4.9796 | 4.9263 | 4.8911 | 4.8661 | 4.8474 | 4.8329 | 4.8213 |
| $k^2 = 0.1$ | 6.7646 | 5.5026 | 5.2488 | 5.1276 | 5.0561 | 5.0090 | 4.9755 | 4.9505 | 4.9311 | 4.9156 |
| $k^2 = 0.2$ | 7.2721 | 5.8164 | 5.4916 | 5.3347 | 5.2418 | 5.1804 | 5.1368 | 5.1042 | 5.0789 | 5.0587 |
| $k^2 = 0.3$ | 7.8165 | 6.1855 | 5.7901 | 5.5963 | 5.4811 | 5.4047 | 5.3503 | 5.3096 | 5.2780 | 5.2528 |
| $k^2 = 0.4$ | 8.4128 | 6.6211 | 6.1562 | 5.9253 | 5.7872 | 5.6952 | 5.6296 | 5.5804 | 5.5422 | 5.5117 |
| $k^2 = 0.5$ | 9.0831 | 7.1407 | 6.6077 | 6.2832 | 6.1782 | 6.0704 | 5.9932 | 5.9354 | 5.8903 | 5.8542 |
| $k^2 = 0.6$ | 9.8641 | 7.7745 | 7.1737 | 6.8680 | 6.6828 | 6.5587 | 6.4698 | 6.4029 | 6.3507 | 6.3090 |
| $k^2 = 0.7$ | 10.824 | 8.5802 | 7.9086 | 7.5634 | 7.3533 | 7.2078 | 7.1105 | 7.0341 | 6.9745 | 6.9266 |
| $k^2 = 0.8$ | 12.113 | 9.6896 | 8.9362 | 8.5456 | 8.3068 | 8.1457 | 8.0298 | 7.9424 | 7.8741 | 7.8194 |
| $k^2 = 0.9$ | 14.215 | 11.526 | 10.654 | 10.199 | 9.9192 | 9.7300 | 9.5936 | 9.4906 | 9.4101 | 9.3454 |
| $k^2 = 1.0$ | — | — | — | — | — | — | — | — | — | — |

Table V. Values of $q_{\min}$ (see eq.22) in units of Schwarzschild radii for various values of $e$ and $k^2$ in Region I.

| $q_{\min}$ | $e = 1$ | $e = 2$ | $e = 3$ | $e = 4$ | $e = 5$ | $e = 6$ | $e = 7$ | $e = 8$ | $e = 9$ | $e = 10$ |
|---|---|---|---|---|---|---|---|---|---|---|
| $k^2 = 0.001$ | 1001.0 | 1333.5 | 1499.7 | 1599.5 | 1666.0 | 1713.5 | 1749.1 | 1776.8 | 1799.0 | 1817.1 |
| $k^2 = 0.01$ | 101.00 | 133.46 | 149.73 | 159.49 | 165.99 | 170.64 | 174.12 | 176.83 | 179.00 | 180.78 |
| $k^2 = 0.1$ | 11.000 | 13.502 | 14.777 | 15.542 | 16.052 | 16.416 | 16.689 | 16.901 | 17.070 | 17.209 |
| $k^2 = 0.2$ | 6.0000 | 6.8785 | 7.3327 | 7.6055 | 7.7871 | 7.9166 | 8.0135 | 8.0887 | 8.1489 | 8.1980 |
| $k^2 = 0.3$ | 4.3333 | 4.6985 | 4.8890 | 5.0035 | 5.0797 | 5.1339 | 5.1744 | 5.2059 | 5.2310 | 5.2515 |
| $k^2 = 0.4$ | 3.5000 | 3.6280 | 3.6951 | 3.7354 | 3.7621 | 3.7811 | 3.7953 | 3.8063 | 3.8151 | 3.8223 |
| $k^2 = 0.5$ | 3.0000 | 3.0000 | 3.0000 | 3.0000 | 3.0000 | 3.0000 | 3.0000 | 3.0000 | 3.0000 | 3.0000 |
| $k^2 = 0.6$ | 2.6667 | 2.5920 | 2.5530 | 2.5298 | 2.5144 | 2.5035 | 2.4954 | 2.4892 | 2.4842 | 2.4801 |
| $k^2 = 0.7$ | 2.4286 | 2.3087 | 2.2466 | 2.2097 | 2.1854 | 2.1682 | 2.1554 | 2.1455 | 2.1377 | 2.1313 |
| $k^2 = 0.8$ | 2.2500 | 2.1024 | 2.0266 | 1.9819 | 1.9525 | 1.9318 | 1.9164 | 1.9045 | 1.8951 | 1.8875 |
| $k^2 = 0.9$ | 2.1111 | 1.9466 | 1.8632 | 1.8142 | 1.7822 | 1.7597 | 1.7430 | 1.7302 | 1.7200 | 1.7117 |
| $k^2 = 1.0$ | 2.0000 | 1.8257 | 1.7384 | 1.6875 | 1.6544 | 1.6311 | 1.6139 | 1.6007 | 1.5902 | 1.5817 |

Table VI. Values of $x_{a_1}$ (see eq.34) in units of Schwarzschild radii for various values of $e$ and $k^2$ in Region I.

| $x_{a_1}$ | $e = 1$ | $e = 2$ | $e = 3$ | $e = 4$ | $e = 5$ | $e = 6$ | $e = 7$ | $e = 8$ | $e = 9$ | $e = 10$ |
|---|---|---|---|---|---|---|---|---|---|---|
| $k^2 = 0.001$ | — | -2668.6 | -2250.7 | -2133.6 | -2083.4 | -2057.0 | -2041.4 | -2031.4 | -2024.6 | -2019.7 |
| $k^2 = 0.01$ | — | -268.58 | -225.75 | -219.81 | -217.21 | -205.58 | -203.92 | -202.84 | -202.09 | -201.56 |
| $k^2 = 0.1$ | — | -28.761 | -23.380 | -21.745 | -20.985 | -20.554 | -20.280 | -20.091 | -19.954 | -19.850 |
| $k^2 = 0.2$ | — | -15.639 | -12.288 | -11.222 | -10.706 | -10.404 | -10.206 | -10.066 | -9.9627 | -9.8830 |
| $k^2 = 0.3$ | — | -11.418 | -8.7089 | -7.8219 | -7.3830 | -7.1213 | -6.9477 | -6.8240 | -6.7316 | -6.6598 |
| $k^2 = 0.4$ | — | -9.4292 | -7.0159 | -6.2110 | -5.8073 | -5.5644 | -5.4021 | -5.2858 | -5.1985 | -5.1305 |
| $k^2 = 0.5$ | — | -8.3382 | -6.0832 | -5.3218 | -4.9369 | -4.7039 | -4.5475 | -4.4353 | -4.3507 | -4.2847 |
| $k^2 = 0.6$ | — | -7.6986 | -5.5346 | -4.7978 | -4.4234 | -4.1960 | -4.0430 | -3.9330 | -3.8500 | -3.7851 |
| $k^2 = 0.7$ | — | -7.3188 | -5.2077 | -4.4851 | -4.1167 | -3.8874 | -3.7415 | -3.6328 | -3.5507 | -3.4865 |
| $k^2 = 0.8$ | — | -7.1021 | -5.0209 | -4.3062 | -3.9412 | -3.7187 | -3.5688 | -3.4607 | -3.3791 | -3.3153 |
| $k^2 = 0.9$ | — | -6.9947 | -4.9282 | -4.2174 | -3.8539 | -3.6324 | -3.4829 | -3.3752 | -3.2939 | -3.2302 |
| $k^2 = 1.0$ | — | -6.9640 | -4.9016 | -4.1919 | -3.8290 | -3.6076 | -3.4583 | -3.3507 | -3.2694 | -3.2059 |

Table VII. Values of $x_{a_2}$ (see eq.34) in units of Schwarzschild radii for various values of $e$ and $k^2$ in Region I.

| $x_{a_2}$ | $e = 1$ | $e = 2$ | $e = 3$ | $e = 4$ | $e = 5$ | $e = 6$ | $e = 7$ | $e = 8$ | $e = 9$ | $e = 10$ |
|---|---|---|---|---|---|---|---|---|---|---|
| $k^2 = 0.001$ | — | -2672.2 | -2252.0 | -2134.3 | -2083.8 | -2057.3 | -2041.6 | -2031.5 | -2024.7 | -2019.8 |
| $k^2 = 0.01$ | — | -272.41 | -227.06 | -220.52 | -217.66 | -205.88 | -204.13 | -203.00 | -202.22 | -201.66 |
| $k^2 = 0.1$ | — | -35.216 | -25.480 | -22.861 | -21.705 | -21.071 | -20.676 | -20.409 | -20.218 | -20.076 |
| $k^2 = 0.2$ | — | -29.821 | -16.086 | -13.205 | -12.006 | -11.360 | -10.961 | -10.691 | -10.498 | -10.352 |
| $k^2 = 0.3$ | — | -100.13 | -17.047 | -11.727 | -9.8837 | -8.9632 | -8.4158 | -8.0545 | -7.7989 | -7.6090 |
| $k^2 = 0.4$ | — | 24.262 | -51.112 | -16.776 | -11.681 | -9.6678 | -8.5987 | -7.9387 | -7.4918 | -7.1697 |
| $k^2 = 0.5$ | — | 9.3864 | 17.546 | 6.1133 X $10^9$ | -44.904 | -21.365 | -15.323 | -12.570 | -11.001 | -9.9882 |
| $k^2 = 0.6$ | — | 6.5652 | 6.5320 | 8.1665 | 10.800 | 14.743 | 20.918 | 31.699 | 54.989 | 141.69 |
| $k^2 = 0.7$ | — | 8.3133 | 4.7770 | 4.3893 | 4.4475 | 4.6487 | 4.8646 | 5.1091 | 5.3546 | 5.5946 |
| $k^2 = 0.8$ | — | -23.034 | 9.7872 | 5.2350 | 4.1473 | 3.6958 | 3.4629 | 3.3276 | 3.2430 | 3.1874 |
| $k^2 = 0.9$ | — | -6.8790 | -4.7820 | -5.6443 | -7.6795 | -11.495 | -19.786 | -49.097 | 214.95 | 39.112 |
| $k^2 = 1.0$ | — | — | — | — | — | — | — | — | — | — |

Table VIII. Representative values for spacecraft deflections.

| Parameter | Earth | Jupiter | Sun |
|---|---|---|---|
| $r_p$ [km] | 6678 | 71700 | $2.784 \times 10^6$ |
| $v_\infty$ [km/s] | 9.000 | 5.455 | 37.92 |
| $GM/c^2$ [km] | $4.435 \times 10^{-6}$ | $1.410 \times 10^{-3}$ | 1.476 |
| $e$ | 2.358 | 1.017 | 1.030 |
| $s$ | $1.406 \times 10^{-5}$ | $9.875 \times 10^{-5}$ | $5.111 \times 10^{-4}$ |
| $180° - 2\phi_0$ [°] | 50.20 | 159.1 | 152.3 |
| $\Delta\phi$ | $4.224 \times 10^{-9}$ | $1.838 \times 10^{-7}$ | $4.925 \times 10^{-6}$ |

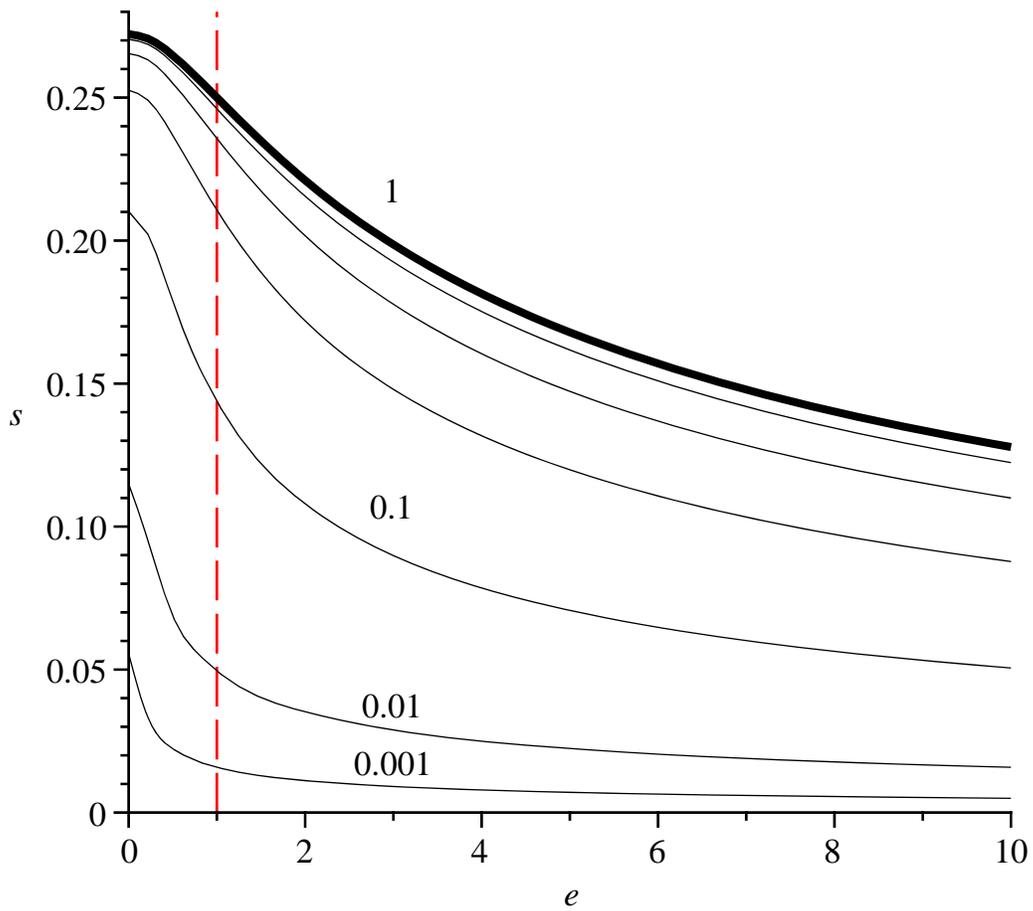

Fig.1 Region I - Plots of constant $k^2$ = 0.001, 0.01, 0.1, 0.3, 0.5, 0.7, 1. The hyperbolic-type trajectories are defined in the sector where e > 1 and also $k^2 \leq 1$.

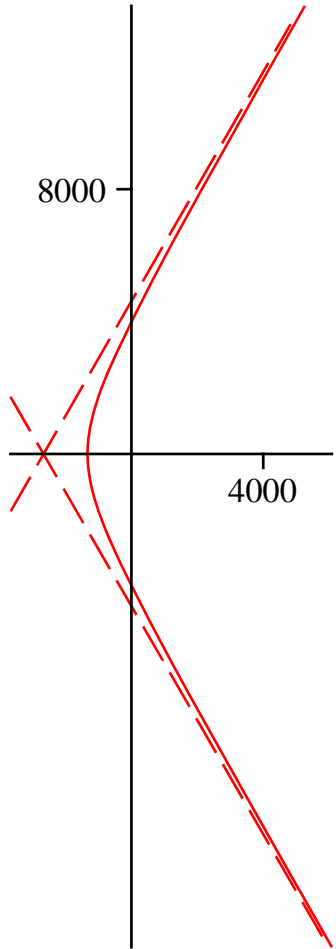

Fig. 2  Hyperbolic-type orbit for e = 2 and $k^2$ = 0.001. The dotted lines are the asymptotes. Distances are in units of Schwarzschild radii.

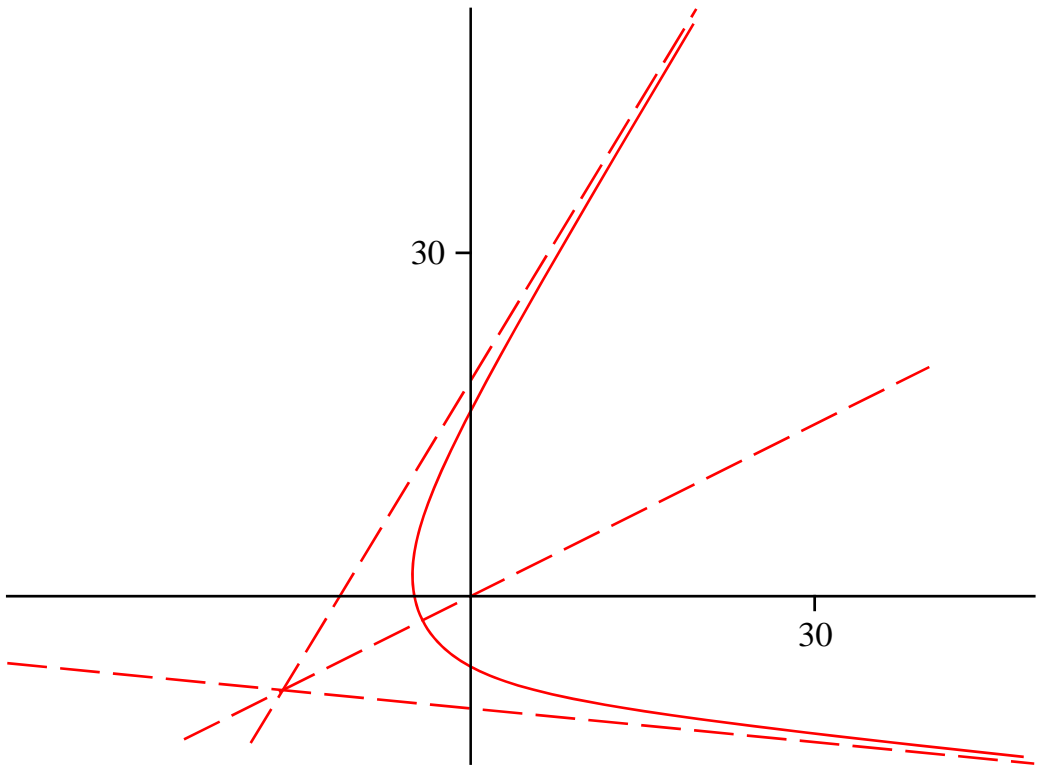

Fig.3 Hyperbolic-type orbit for e = 2 and $k^2$ = 0.3. The dotted lines are the asymptotes and the axis of symmetry. Distances are in units of Schwarzschild radii.

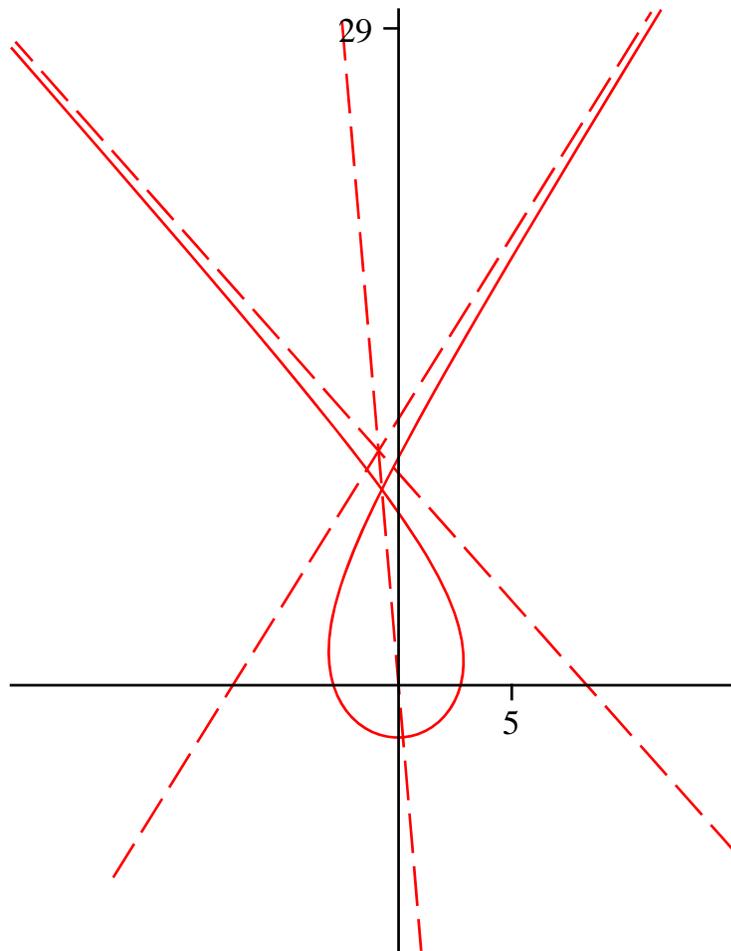

Fig.4  Hyperbolic-type orbit for e = 2 and $k^2$ = 0.7. Distances are in units of Schwarzschild radii.

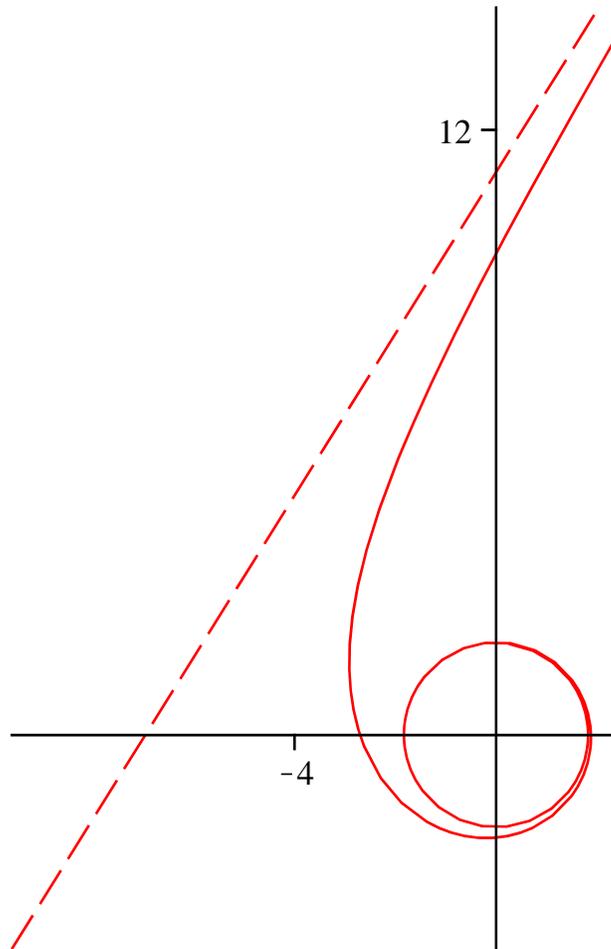

Fig. 5 Asymptotic hyperbolic-type orbit for e = 2 and $k^2 = 1$. The dashed line is an asymptote. Distances are in units of Schwarzschild radii.

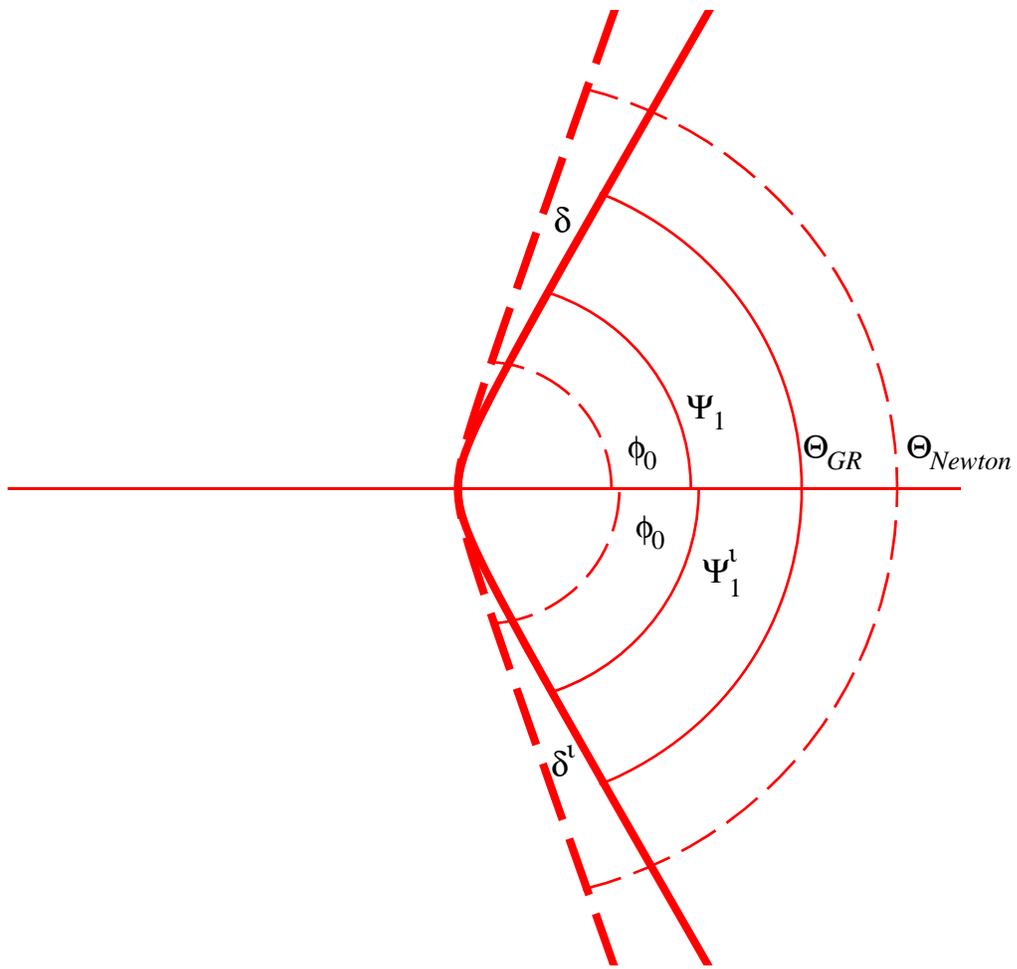

Fig.6 Illustration of the angles defined at the beginning of Section 4 on small deviations from Newtonian trajectories.